\documentclass[runningheads]{llncs}
\usepackage{graphicx}

\usepackage{booktabs} 
\usepackage{multirow}
\usepackage{multicol}
\usepackage[font=small,labelfont=bf]{caption} 
\usepackage{balance}
\usepackage{comment}
\usepackage{amsmath}
\usepackage{subcaption}
\usepackage{wrapfig}
\usepackage{etoolbox}
\usepackage{mathrsfs}
\usepackage{mathtools}
\begin{document}
\abovedisplayskip=0pt
\abovedisplayshortskip=0pt
\belowdisplayskip=0.5pt
\belowdisplayshortskip=0pt
\makeatletter
\let\origsection\section
\renewcommand\section{\@ifstar{\starsection}{\nostarsection}}

\newcommand\nostarsection[1]
{\sectionprelude\origsection{#1}\sectionpostlude}

\newcommand\starsection[1]
{\sectionprelude\origsection*{#1}\sectionpostlude}

\newcommand\sectionprelude{%
	\vspace{-1.05em}
}

\newcommand\sectionpostlude{%
	\vspace{-0.7em}
}

\let\origsubsection\subsection
\renewcommand\subsection{\@ifstar{\starsubsection}{\nostarsubsection}}

\newcommand\nostarsubsection[1]
{\subsectionprelude\origsubsection{#1}\subsectionpostlude}

\newcommand\starsubsection[1]
{\subsectionprelude\origsubsection*{#1}\subsectionpostlude}

\newcommand\subsectionprelude{%
	\vspace{-1.05em}
}

\newcommand\subsectionpostlude{%
	\vspace{-0.7em}
}

\makeatletter
\renewcommand\footnotesize{%
	\@setfontsize\footnotesize\@ixpt{1}%
	\abovedisplayskip 8\p@ \@plus2\p@ \@minus4\p@
	\abovedisplayshortskip \z@ \@plus\p@
	\belowdisplayshortskip 5\p@ \@plus2\p@ \@minus2\p@
	\def\@listi{\leftmargin\leftmargini
		\topsep 4\p@ \@plus2\p@ \@minus2\p@
		\parsep 2\p@ \@plus\p@ \@minus\p@
		\itemsep \parsep}%
	\belowdisplayskip \abovedisplayskip
}

\makeatother	

\title{Iterative Relevance Feedback for Answer Passage Retrieval with Passage-level Semantic Match}
\titlerunning{IRF for Answer Passage Retrieval with Passage-level Semantic Match}
%
 \author{Keping Bi  \and
	Qingyao Ai  \and
	W. Bruce Croft }
%
%
 \institute{College of Information and Computer Sciences, University of Massachusetts Amherst, Amherst, MA, USA \\
 	\email{\{kbi,aiqy,croft\}@cs.umass.edu}}
\maketitle              

\begin{abstract}
	Relevance feedback techniques assume that users provide relevance judgments for the top k (usually 10) documents and then re-rank using a new query model based on those judgments. Even though this is effective, there has been little research recently on this topic because requiring users to provide substantial feedback on a result list is impractical in a typical web search scenario. 
	In new environments such as voice-based search with smart home devices, however, feedback about result quality can potentially be obtained during users' interactions with the system. 
	Since there are severe limitations on the length and number of results that can be presented in a single interaction in this environment, the focus should move from browsing result lists to iterative retrieval and from retrieving documents to retrieving answers. 
	In this paper, we study iterative relevance feedback techniques with a focus on retrieving answer passages. 
	We first show that iterative feedback is more effective than the top-k approach for answer retrieval. 
	Then we propose an iterative feedback model based on passage-level semantic match and show that it can produce significant improvements compared to both word-based iterative feedback models and those based on term-level semantic similarity.
	\keywords{Iterative Relevance Feedback; Answer Passage Retrieval; Passage Embeddings}
\end{abstract}

\section{Introduction}
\label{sec:introduction}
In typical relevance feedback (RF) techniques, users are provided with a list of top-ranked documents and asked to assess their relevance. The judged documents, together with the original query, are used to estimate a new query model using an RF model, which further acts as a basis for re-ranking.  
There were extensive studies of RF \cite{aalbersberg1992incremental,brondwine2016utilizing,dehghani2016luhn,rocchio1971relevance,robertson1976relevance,salton90improvingretrieval,lwayama2000relevance,allan1996incremental,ruthven2003survey,lavrenko2001relevance,zhai2001model} based on the vector space model (VSM) \cite{salton1975vector}, the probabilistic model \cite{maron1960relevance} and, more recently, on the language model (LM) for Information Retrieval (IR) approach \cite{ponte1998language}. Despite the effectiveness of RF, the overhead involved in obtaining user relevance judgments has meant that it is not used in typical search scenarios. 

With mobile and voice-based search becoming more popular, it becomes feasible to obtain feedback about result quality during users' interactions with the system. 
In these scenarios, the display space or voice bandwidth leads to severe limitations on the length and number of results shown in a single interaction. 
Thus, instead of providing a list of results, an iterative approach to feedback may be more effective. 
There has been some work in the past on iterative relevance feedback (IRF) with only a few results in each interaction using the VSM \cite{aalbersberg1992incremental,lwayama2000relevance,allan1996incremental}, but this has not been looked at for a long time. In addition, the space and bandwidth limitations make the retrieval of longer documents less desirable than shorter answer passages. Motivated by these reasons, in this paper, we present a detailed study of methods for IRF focused on answer passage retrieval.

Although they could be applied to any text retrieval scenario, most existing RF algorithms use word-based models originally designed for document retrieval. Answer passages, however, are much shorter than documents, which could potentially present problems for accurate estimation of word weights in the existing word-based RF methods. Moreover, the limitations on the length and number of results in IRF mean that there is even less relevant text available at every iteration. Given these issues, introducing complementary information from semantic space may help to estimate a more accurate RF model. 
Dense vector representations of words and paragraphs in distributed semantic space, called embeddings, \cite{mikolov2013distributed,le2014distributed,sun2015learning,dai2015document,chen2017efficient}, have been effectively applied to many natural language processing (NLP) tasks.
Embeddings have also been used in pseudo relevance feedback based on documents \cite{Zamani:2016:EQL:2970398.2970405,Rekabsaz:2016:GTM:2983323.2983833}, but their impact in iterative and passage-based feedback is not known. 
Besides, these previous work use semantic similarity at the term level and does not consider semantic match at larger granularity.
This had led us to incorporate passage-level semantic match in IRF for answer passage retrieval to improve upon word-based IRF and other embedding-based IRF using term-level semantic similarity. 


In the paper, we first investigate 
whether iterative feedback based on different frameworks is effective relative to RF
with a list of top k (k=10) results on answer passage retrieval.
The results indicate that IRF is significantly more effective on answer passage collections.
In addition, we propose an embedding-based IRF method using passage-level similarity for answer passage retrieval. This method incorporates the similarity scores computed with different types of answer passage embeddings and fuses them with other types of IRF models. 
The model we propose significantly outperforms IRF baselines based on words or semantic matches between terms. Combining both term-level and passage-level semantic match information leads to additional gains in performance.

\section{Related Work}
\label{sec:related_work}
In this section, we first review previous approaches to RF and IRF. We then discuss related work on embeddings of words and paragraphs applied to IR and some previous studies on answer passage retrieval.

\textbf{Relevance Feedback.}
In general, there are mainly three types of relevance feedback methods for ad-hoc retrieval, which are based on the vector space model (VSM) \cite{salton1975vector}, the probabilistic model \cite{maron1960relevance} and the language model (LM) \cite{ponte1998language}. Basically, they all extract expansion terms from annotated relevant documents and re-weight the original query terms so as to estimate a more accurate query model to retrieve better results.

Rocchio \cite{rocchio1971relevance} is generally credited as the first RF technique, developed on the VSM. It refines the vector of a user query by bringing it closer to the average vector of relevant documents and further from the average vector of non-relevant documents. In the probabilistic model, expansion terms are scored according to the probability they occur in relevant documents compared to non-relevant documents \cite{robertson1976relevance,harman1992relevance}. 
Salton et al. \cite{salton90improvingretrieval} studied various RF techniques based on the VSM and probabilistic model and showed
that the probabilistic RF models are in general not as effective as the methods in the VSM. 

More recently, feedback techniques have been investigated extensively based on LM, among which, the relevance model \cite{lavrenko2001relevance} and the mixture model \cite{zhai2001model} are two well-known examples that empirically perform well. 
In the third version of the relevance model (RM3) \cite{lavrenko2001relevance}, 
the probabilities of expansion terms are estimated with occurrences of the terms in feedback documents. The mixture model \cite{zhai2001model} considers a feedback document to be generated from a mixture of a corpus language model, and a query topic model, which is estimated with the EM algorithm. 
Some recent work \cite{brondwine2016utilizing,dehghani2016luhn} extend the mixture model by considering additional or different language models as components of the mixture.  

\textbf{Iterative Relevance Feedback.}
In contrast to most RF systems that ask users to give relevance assessments on a batch of documents, 
Aalsberg et al. \cite{aalbersberg1992incremental} proposed the alternative technique of incremental RF based on Rocchio. 
Users are asked to judge a single result shown in each interaction, then the query model can be modified iteratively through feedback. This approach showed higher retrieval quality compared with standard batch feedback. Later, Lwayama et al. \cite{lwayama2000relevance} showed that the incremental relevance feedback used by Aalsberg et al. works better for documents with similar topics, while not as well for documents spanning several topics. 
In this paper, we investigate how IRF performs on retrieval of answer passages instead of documents using more recent retrieval models.

Some recent TREC tracks \cite{yang2016trec,grossman2016trec} have made use of iterative and passage-level feedback, but they focus on document retrieval with different objectives and require a large amount of user feedback.
The Total Recall track  \cite{yang2016trec} aims at high recall, where the goal is to promote all of the relevant documents before non-relevant ones.
The target of the Dynamic Domain track \cite{grossman2016trec} is to identify documents satisfying all the aspects of the users' information need with passage-level feedback. 
In contrast, we focus on iterative feedback for the task of answer passage retrieval and investigate IRF with a fixed small amount of feedback. 

\textbf{Word and Paragraph Embeddings for RF.}
Dense representations, called embeddings, of words and paragraphs, have become popular and been used \cite{mikolov2013distributed,zamani:2017:RWE,sun2015learning,chen2017efficient} to abstract the meaning from a piece of text in semantic space. Two well-known techniques to train word and paragraph embeddings are Word2Vec  \cite{mikolov2013efficient} and Paragraph Vectors (PV) \cite{le2014distributed} respectively.
The similarity of word embeddings can be used to compute the transition probabilities between words \cite{Rekabsaz:2016:GTM:2983323.2983833,Zamani:2016:EQL:2970398.2970405} and incorporated with the VSM or relevance model \cite{lavrenko2001relevance} to solve problems of term mismatch. 
Basically, these approaches use semantic match at word level and are in the form of query expansion. In contrast, our approach uses semantic match at passage level and is not based on query expansion.

\section{Word-based IRF Models}
\label{sec:iter_model}
In IRF, the topic model of users' intent can be refined each iteration after a small number of results are assessed. 
Therefore, re-ranking is triggered earlier in IRF than in standard top-k RF methods.
On the one hand, earlier re-ranking may produce better results with fewer iterations, which essentially reduces the user efforts in search interactions. On the other hand, having only a small amount of feedback information in each iteration may hurt the accuracy of model estimation and cause topic drift in the iterative process.  

We convert several representative models to iterative versions and investigate the performance of the IRF models on answer passage retrieval. 
Since LM and VSM are the two most effective frameworks for RF, 
we study iterative feedback under these two frameworks. 
We use RM3 \cite{lavrenko2001relevance} and the Distillation (or Distill) model \cite{brondwine2016utilizing} to represent the LM framework, and Rocchio \cite{rocchio1971relevance} for VSM.
RM3 is a common baseline for pseudo RF that has also been used for RF. 
Distillation is one of the most recent RF methods, which is an extension of the mixture model by incorporating a query specific non-relevant topic model. Rocchio \cite{rocchio1971relevance} is the standard feedback model in VSM.
As for the retrieval models for initial ranking, we use Query Likelihood (QL) for LM, 
and BM25 \cite{robertson1995okapi} for VSM respectively. 

To keep the query model from diverging to non-relevant topics, we maintain two pools for relevant and non-relevant results, which accumulate all the judgments until the $i$th iteration.
During the $i$th iteration, judged relevant results and non-relevant results are added to the corresponding pool. Expanded query models are then estimated from the relevant document pool by RM3 and from both relevant and non-relevant pools by Distillation and Rocchio.
Detailed introduction about the IRF models and the experiments can be found in \cite{bi2018revisiting}.

\section{Passage Embedding based IRF Models}
\label{sec:pv_irf_answer}
Word-based RF methods were initially designed for document retrieval and usually based on query expansion. In contrast to documents, answer passages do not have sufficient text to estimate the probabilities or weights of the expansion terms accurately, especially for IRF when fewer results are available in each iteration. 
To alleviate the problem of text insufficiency in IRF, we incorporate semantic information about paragraphs to the IRF models. Paragraph embeddings are shown to be capable of capturing the semantic meanings of passages \cite{le2014distributed,sun2015learning,chen2017efficient}, which could potentially help us build more robust IRF models by supporting semantic matching between passages.

In this section, we propose to use paragraph embeddings to improve the performance of IRF for answer passage retrieval.
In contrast to existing word-based and embedding-based RF methods, this approach does not extract expansion terms to update the query model.
Instead, it represents the relevance topic from feedback passages with embeddings. 
Similar to Rocchio, we assume a relevant passage should be near the centroid of other relevant passages in the embedding space. 
Also, we only focus on positive feedback as negative feedback has been shown to have little benefit for RF when positive feedback is available in previous studies \cite{aalbersberg1992incremental}.
Therefore, our model can be viewed as an embedding version of Rocchio with only positive feedback.

We first describe the methods we use to obtain the semantic representations for answer passages.
Then we will introduce the passage embedding based iterative feedback model. 
\subsection{Passage Embeddings} 
\textbf{Aggregated Word Embeddings.}
One common way of representing passages is to use aggregated embeddings of words in the paragraph.
Word2Vec is a well-known method of training word embeddings \cite{mikolov2013efficient,mikolov2013distributed}. It projects words to dense vector space and uses a word to predict its context or predicts a word by its context. 
In our experiments, we also use average word embeddings trained from Word2Vec both with and without IDF weighting as passage representations. 

\begin{wrapfigure}[13]{r}{0.55\textwidth}
	\centering
	\vspace{-20pt}
	\includegraphics[width=2.6in]{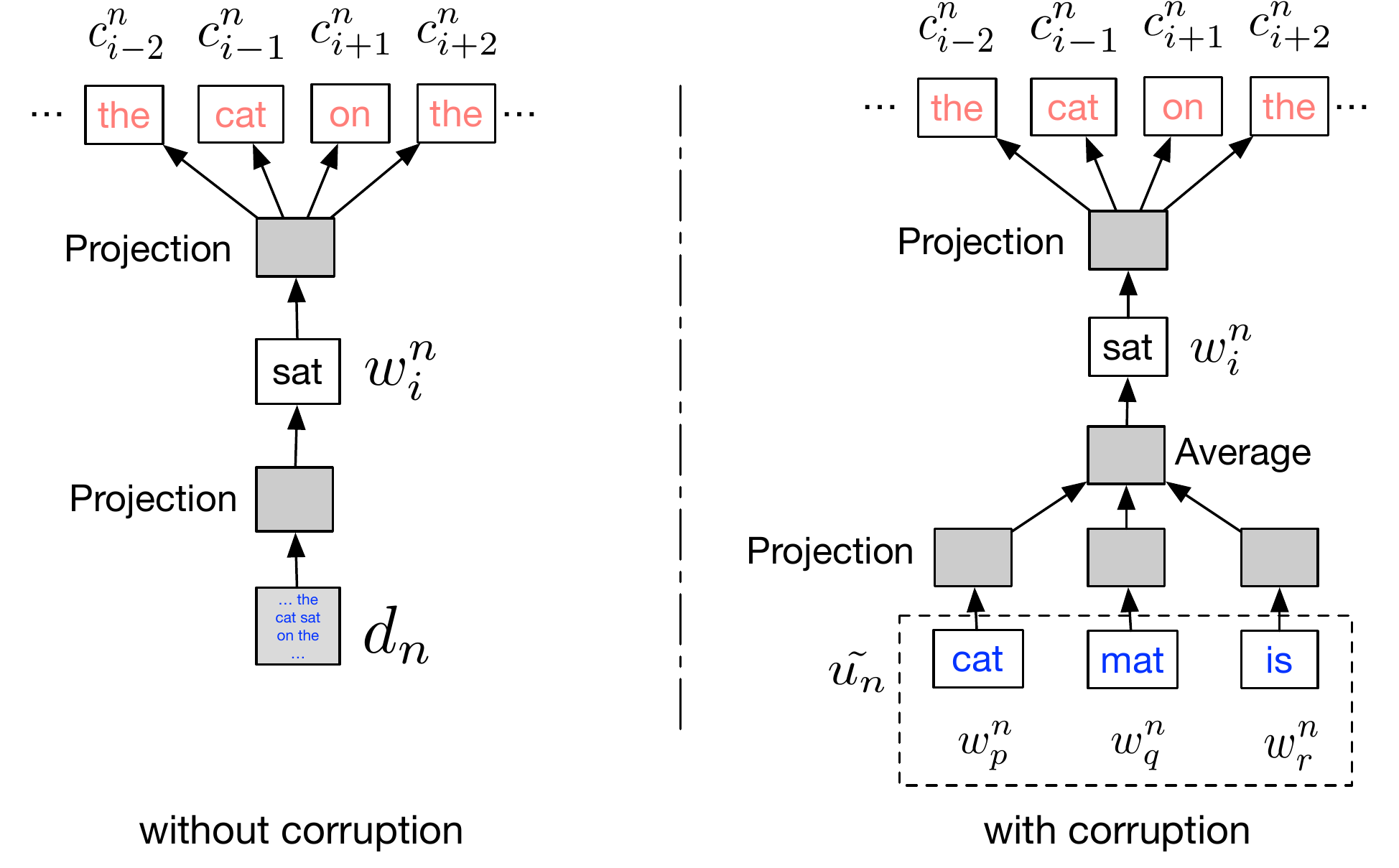}
	\caption{HDC models used in our experiments. Red words are local context, and blue words are global context.}
	\label{fig:hdc_model}
\end{wrapfigure}

\textbf{Paragraph Vectors.}
\label{sec:hdc_model}
The other way of representing passages is using specially designed paragraph vectors models
as in \cite{le2014distributed,chen2017efficient,sun2015learning}. The models we use are PV-HDC \cite{sun2015learning} with or without corruption, shown in Figure \ref{fig:hdc_model}. 
PV-HDC is an extension of the initially proposed paragraph vector model \cite{le2014distributed}, where a document vector is first used to predict an observed word, and afterward, the observed word is used to predict its context words.
The recent work of training paragraph representation through corruption \cite{chen2017efficient} shows advantages in many tasks such as sentiment analysis. 
It replaces the original part of paragraph representation with a corruption module, where the global context $\tilde{u}$ is generated through an unbiased dropout corruption at each update and the paragraph representation is calculated as the average embeddings of the words in $\tilde{u}$. The final representation is simply the average of the embeddings of all the words in the paragraph. 
We also investigate other models such as the original PV models, DM, DBOW, \cite{le2014distributed}, and the Parallel Document Context Model (PDC) \cite{sun2015learning}, both with and without corruption, but HDC is better in most cases.  So we exclude the other models in the paper. 

\subsection{IRF with Passage-level Semantic Similarity}
As an alternative to query expansion based RF methods, we propose to represent the whole semantic meaning of a passage and a passage set with vectors in the embedding space and measure the similarity between them without explicitly extracting any expansion terms.
Specifically, we represent the relevance topic in the $i$th iteration as the embedding of the relevant passage pool and fuse the similarity between a passage with the relevance topic with other RF methods.  
Thus the score function is shown as follows, 
\begin{equation}
\label{eq:sf_total}
score(Q^{(i)}, d) = score_{rf}(Q^{(i)}, d) + \lambda_{sf} score_{sem}(RP^{(i)}, d) \\
\end{equation}
$Q^{(i)}$ is the expanded query model estimated by iterative version of RF models such as RM3, Distillation and Rocchio; $d$ is the candidate passage;  
$RP^{(i)}$ denotes the relevant passage pool in the $i$th iteration; 
$score_{rf}$ denotes the score calculated from other RF models; $score_{sem}$ is the semantic match score between passages, which is the commonly used cosine similarity in the paper;
$\lambda_{sf}$ is the coefficient of incorporating the passage embedding based similarity;
Similar to Rocchio, we assume the topic of a passage set is the centroid of these passages
and we consider a relevant passage pool can be represented by average vectors of the passages in it.
Thus the similarity between a passage and the pool is 
\begin{equation}
\label{eq:w_cosine}
score_{sem}(RP^{(i)}, d) = \cos(\frac{1}{|RP^{(i)}|}\!\!\!\!\sum_{~~~~d_r\in RP^{(i)}}{\!\!\!\!\!\!\vec{d_r}}, ~~~~\vec{d}) \\
\end{equation}
where $\vec{d_r}$ and $\vec{d}$ is the vector representation of $d_{r}$ and $d$ in the embedding space.

Our method has two advantages over existing RF methods. One is that compared to expansion term based methods that only alleviate word-level mismatch, semantic similarity of larger granularity is captured in our method. The other is the flexibility of combining this semantic match signal with different types of approaches such as RM3, Distillation, Mixture, Rocchio, and other embedding-based feedback approaches.



\section{Experiments of Word-based IRF} 
\label{sec:irf_exp_doc_psg}
In this section, we introduce the experimental setup and results of word-based IRF on answer passage retrieval. 
\subsection{Experimental Setup}\label{sec:IRF_metrics}

\textbf{Data.}
In our experiments, we used WebAP and PsgRobust for answer passage retrieval. Statistics of the datasets are summarized in Table~\ref{tab:dataset}.
WebAP~\cite{Yang2016BeyondFQ} is a web answer passage collection built on Gov2. 
It uses a subset of queries that are likely to have passage-level answers from Gov2 and retrieved the top 50 documents with the Sequential Dependency Model (SDM)~\cite{metzler2005markov}. 
After that, relevant documents were annotated for relevant answer passages. Overall, 3843 passages from 1200 documents are annotated as relevant.  
In our experiments, we split the rest of the documents into non-overlapping 2 or 3 (randomly chosen) contiguous sentences as non-relevant passages and used topic descriptions as questions.

PsgRobust \footnote{This dataset is publicly available at \url{https://ciir.cs.umass.edu/downloads/PsgRobust/}.}
is a new collection we created for answer passage retrieval.
It is based on the TREC Robust collection following a similar approach as WebAP but without manual annotation.
In PsgRobust, we assume that \textit{top-ranked passages in relevant documents can be considered as relevant} and \textit{all passages in non-relevant documents are irrelevant}.
We first retrieved the top 100 documents for each title query in Robust with SDM~\cite{metzler2005markov} and generated answer passages from them with a sliding window of random lengths (2 or 3 sentences) and no overlap.
After that, we retrieved top 100 passages with SDM again and treated those from relevant documents as the relevant passages.
Similar to WebAP, we used the descriptions of Robust topics as questions and have 246 queries with non-zero relevant answer passages in total.
The Recall@100 in the initial retrieval process is 0.43, which means that 43\% of relevant documents for all queries were included in the passage collection on average.
By manually checking some randomly sampled passages marked as relevant, we found most of them are indeed relevant passages for the questions. There are 6589 relevant passages from 3544 documents for the 246 queries in total.

We also considered other collections that have passage-level annotation such as the DIP2016Corpus 
\cite{habernal2016new} and the dataset from the Dynamic Domain track \cite{yang2016trec}.
However, they either are trivial for RF tasks (almost all top 10 results retrieved by BM25 are relevant in DIP2016Corpus) or have few queries (only 26 and 27 for the two domains of the Dynamic Domain track). 
Other popular question answering datasets usually only have one relevant answer for each query and thus are not suitable for our RF task either.
Therefore, we only report the results of WebAP and PsgRobust in this paper.

\begin{table} [t]
	\centering
	\caption{Statistics of experimental datasets.}
	\label{tab:dataset}
	\begin{tabular}{ l || r |  r  | r | r | r | r | r } 
		\hline
		Dataset & \#Psg & PsgLen & Vocab & \#Query & \#Psg/D & \#RelPsg/Q & \#RelPsg/D\\ \hline
		WebAP & 379k & 45 & 59k & 80 & 114.3 & 48.0 & 3.2 \\ 
		PsgRobust & 383k & 46 & 64k & 246 & 17.1 & 26.8 & 1.9 \\ 
		\hline
	\end{tabular}
\end{table}
\textbf{System Settings}.
All the methods we implemented are based on the Galago toolkit \cite{croft2010search} \footnote{\url{http://www.lemurproject.org/galago.php}}. 
Stopwords were removed from all collections using the standard INQUERY stopword list and words were stemmed with Krovetz Stemmer \cite{krovetz1993viewing}. 
To compare iterative feedback with typical top-k feedback in a fair manner, we fixed the total number of judged results as 10 and experimented with 1, 2, 5, and 10 iterations, where 10, 5, 2, 1 results were judged during each iteration respectively. Then 10Doc-1Iter is exactly the top-k feedback. 
We pay more attention to the settings of one or two results per iteration which are more likely to be in a real interactive search scenario considering the limitation of presenting results. 
True labels of results were used to simulate users' judgments. 

All the parameters were set using 5-fold cross-validation over all the queries in each collection with grid search.
For WebAP and PsgRobust, we tuned $\mu$ of QL in $\{30,50,300,500,1000,1500\}$ and $k$ of BM25 from $\{1.2,1.4,\cdots,2\}$, b set to 0.75 as suggested by \cite{mogotsi2010christopher}. The number of expansion terms $m$ is from $\{10, 20, \cdots, 50 \}$. The range to scan parameters for RM3, Distillation and Rocchio is similar as the corresponding original paper. They are not shown here due to space limits.

\textbf{Evaluation.}
The evaluation should only focus on the ranking of unseen results. So we use freezing ranking \cite{cirillo1969evaluation,ruthven2003survey}, as in \cite{aalbersberg1992incremental,jones2000incremental}, to evaluate the performance of IRF.
The \textsl{freezing ranking} paradigm freezes the ranks of all results presented to the user in the earlier feedback iterations and assigns the first result retrieved in the $i$th iteration rank $iN+1$, where $N$ is the number of results shown in each iteration. 
Note that all the previously shown results are filtered out in the following retrieval to remove duplicates and 
the final result list concatenates $(\#Iter-1)*N$ freezing results with the rest candidates ranked in the last iteration, where $\#Iter$ is the total number of iterations.
Then we use mean average precision at cutoff 100 ($MAP$) and $NDCG@20$ to measure the performance of results overall and on the top. As suggested by Smucker et al. \cite{smucker2007comparison}, statistical significance is calculated with Fisher randomization test with threshold 0.05. 

\subsection{Results and Discussion}
\label{sec:exp_result_IRF_doc_answer}
The performance of the initial retrieval with QL and BM25 and the IRF experimental results are shown in Table \ref{tab:word_irf_doc_psg}. 
All the feedback methods are significantly better than their retrieval baselines, i.e. RM3 and Distillation compared with QL, Rocchio compared with BM25, in terms of both $MAP$ and $NDCG@20$.
\footnote{\label{fn:vsm_worse} On PsgRobust, BM25 and Rocchio underperform QL, RM3 and Distillation respectively by a large margin. Because its labels are generated based on retrieval with SDM, this collection favors approaches in the framework of LM more than VSM.}

In addition, 
on both WebAP and PsgRobust, the $MAP$ and $NDCG@20$ of RM3, Distillation and Rocchio increase as the ten results are judged in more iterations. In other words, IRF is much more effective for answer passage retrieval compared with top-k feedback. Performance goes up when re-ranking is done earlier even when we have only a small number of passages, probably because answer passages are usually focused on a single topic and less likely to cause topic drift. 
Since $MAP$ and $NDCG@20$ show similar trends using IRF under different settings, we only show $MAP$ in Section \ref{sec:exp_IRF_EMB} due to the space limitations. 

\begin{table} [t]
	\caption{Performance of IRF on answer passage collections. $D \times I$ stands for $Doc \times Iter$.  `*' denotes significant improvements over the standard top 10 feedback model ($10\times1$).}
	\label{tab:word_irf_doc_psg}
	\centering
	\scalebox{0.9}{        
		\begin{tabular}{ c | l || l || l | l | l | l || l || l | l | l | l }
			\hline                                        
			\multirow{2}{*}{Dataset} & Method & \multicolumn{5}{c||}{$MAP$ of freezing rank lists}  & \multicolumn{5}{c}{$NDCG@20$ of freezing rank lists} \\            
			\cline{3-12}
			& (D$\times$I) & Initial & (10$\times$1) & (5$\times$2) & (2$\times$5) & (1$\times$10) & Initial & (10$\times$1) & (5$\times$2) & (2$\times$5) & (1$\times$10) \\
			\hline                                        
			\multirow{3}{*}{WebAP}                                        
			& RM3 & 0.076 & 0.100 & 0.107*  & \textbf{0.113*}  & \textbf{0.113*}  & 0.143 & 0.170 & 0.180* & 0.185* & \textbf{0.187*} \\ \cline{2-12}
			& Distill & 0.076 & 0.099 & 0.104$^*$ & 0.109$^*$ & \textbf{0.111}$^*$ & 0.143 & 0.166 & 0.177* & 0.185* & \textbf{0.187}* \\  \cline{2-12}
			& Rocchio & 0.081 & 0.106 &    0.112$^*$ & 0.118$^*$ & \textbf{0.119}$^*$ & 0.150 & 0.169 & 0.181* & 0.190* & \textbf{0.191}* \\
			\hline                
			\multirow{3}{*}{PsgRobust}                                        
			& RM3 & 0.248 & 0.293 & 0.299*  & 0.306*  & \textbf{0.308*} & 0.319 & 0.356 & 0.363* & 0.372* & \textbf{0.373*} \\ \cline{2-12}
			& Distill & 0.248 & 0.292 & 0.299$^*$ & 0.311$^*$ & \textbf{0.313}$^*$ & 0.319 & 0.354    & 0.362* & 0.375* & \textbf{0.379}* \\  \cline{2-12}
			& Rocchio & 0.191 & 0.268 & 0.280$^*$ & 0.285$^*$ & \textbf{0.286}$^*$ & 0.292 & 0.341 & 0.356* & 0.361* & \textbf{0.364}* \\
			\hline    
		\end{tabular}
	}
\end{table}


\section{Experiments of Passage Embedding Based IRF} 
\label{sec:exp_emb_irf_answer}
We compare our method with word-based and embedding-based RF baselines in two groups of experiments.
One is the same as in Section \ref{sec:irf_exp_doc_psg}, i.e. retrieval with a different number of iterations and 10 results judged in total. 
The other focuses on identifying more relevant passages given only one relevant answer passage. 
We first describe the experimental setup and then introduce the two groups of experiments in Section \ref{sec:exp_IRF_EMB} and Section \ref{sec:exp_retrieve_given_onepsg}. 

\subsection{Experimental Setup}
In this part, we again use WebAP and PsgRobust for experiments. 
All comparisons are based on LM (RM3, Distillation, and Rocchio) and VSM (Rocchio) to see whether the complementary semantic match benefits in both frameworks.
We also include the Embedding-based Relevance Model (ERM) \cite{Zamani:2016:EQL:2970398.2970405} as a baseline. 
ERM revises $P(Q|D)$ in the original RM3 as a linear combination of $P(Q|D)$ computed from exact term match and  $P(Q|w,D)$, which takes the semantic relationship between words into account. The translation probability between words is computed with the cosine similarity of their embeddings transformed with the sigmoid function.
\footnote{We also tried the true RF version of BM25-PRF-GT \cite{Rekabsaz:2016:GTM:2983323.2983833}, which is a generalized translation model of BM25 based on word embeddings and Rocchio. Due to its inferior performance on our dataset, we did not include the experiments here.
}
Statistical significance in all the result tables is calculated with Fisher randomization test with threshold 0.05. 

\textbf{Embeddings Training}.
Four paragraph representations are tested in the four groups of experiments, where the base models ($BM$) can be RM3, ERM, Distillation (or Distill) and Rocchio: 

$BM+W2V$/$BM+idfW2V$: uniformly or idf-weighted average word vectors trained with the skip-gram model~\cite{mikolov2013efficient}.  

$BM+PVC$/$BM+PV$: paragraph vectors trained with the HDC structure with or without corruption \cite{sun2015learning,chen2017efficient}.

Embeddings of words or paragraphs were trained with each local corpus respectively. 
Words with the frequency less than 5 were removed. 
No stemming was done across the collections. 
10 negative samples were used for each target word. 
The learning rate and batch size were 0.05 and 256. 
The dimension of embedding vectors was set to 100. 
We also tried other hyper-parameters for training embeddings, and the results were similar under different settings. 
For PVC, corruption rate $q$ \cite{chen2017efficient} was set to 0.9. 
All the neural networks of training embeddings were implemented using TensorFlow \footnote{https://www.tensorflow.org/}. 

\textbf{Parameter Settings.}
We used the best settings of the baseline models and tuned the parameters of the semantic signals with 5-fold cross-validation for different paragraph embeddings. 
All the parameters of ERM are tuned in the same range as \cite{Zamani:2016:EQL:2970398.2970405} suggests.
$\lambda_{sf}$ in equation \ref{eq:sf_total} is selected from $\{5, 10, 15, \cdots, 40 \}$ for WebAP, and $\{0.5, 1, 1.5, \cdots, 5\}$ for PsgRobust respectively. 
\subsection{Iterative Feedback with Embeddings}
\label{sec:exp_IRF_EMB}
First, we conducted IRF experiments with different number of iterations and 10 results judged in total, as described in Section \ref{sec:irf_exp_doc_psg}. 
We use $MAP$ at cutoff 100 of freezing rank lists as the evaluation metric, which is described in section \ref{sec:IRF_metrics}. 
\begin{table} [t]
	\caption{Performance of different IRF models.
		`$*$' and `$\dagger$' denote significant improvements over word-based (RM3, Distillation) and embedding-based (ERM) baselines respectively. ($10Doc \times 1Iter$) represents standard top-k feedback. }
	\centering
	\label{tab:embed_irf_psg}
	\scalebox{0.9}{
		\begin{tabular}{ l || l | l | l | l || l | l | l | l }
			\hline
			Method & \multicolumn{4}{c||}{$MAP$ on \textbf{WebAP}} & \multicolumn{4}{c}{$MAP$ on \textbf{PsgRobust}}  \\
			\cline{2-9}
			(Doc$\times$Iter) & (10$\times$1) & (5$\times$2) & (2$\times$5) & (1$\times$10) & (10$\times$1) & (5$\times$2) & (2$\times$5) & (1$\times$10) \\
			\hline
			RM3  & 0.100  & 0.107  & 0.113  & 0.113  & 0.293  & 0.299  & 0.306  & 0.308 \\
			ERM  & 0.101  & 0.107  & 0.113  & 0.116  & 0.294  & 0.301  & 0.310  & 0.310  \\
			\hline
			RM3+W2V  & \textbf{0.107$^{*\dagger}$}  & \textbf{0.115$^{*\dagger}$}  & 0.117  & 0.116  & \textbf{0.298$^{*\dagger}$}  & 0.303$^{*\dagger}$  & 0.312$^{*\dagger}$  & 0.312$^*$ \\
			RM3+idfW2V  & 0.106$^{*\dagger}$  & 0.113$^{*\dagger}$  & 0.121$^{*\dagger}$  & 0.119$^*$  & \textbf{0.298$^{*\dagger}$}  & 0.303$^*$  & \textbf{0.313$^{*\dagger}$}  & 0.313$^{*\dagger}$\\
			RM3+PV  & 0.102$^*$  & 0.113$^{*\dagger}$  & \textbf{0.123$^{*\dagger}$}  & \textbf{0.123$^*$}  & \textbf{0.298$^{*\dagger}$}  & \textbf{0.305$^{*\dagger}$}  & \textbf{0.313$^*$}  & \textbf{0.314$^*$}\\
			RM3+PVC  & \textbf{0.107$^{*\dagger}$}  & 0.114$^{*\dagger}$  & 0.120$^{*\dagger}$  & 0.114  & 0.297$^{*\dagger}$  & 0.303$^*$  & 0.308  & 0.311$^*$ \\
			\hline
			ERM+W2V  & \textbf{0.107$^{*\dagger}$}  & \textbf{0.116$^{*\dagger}$}  & 0.119$^{*\dagger}$  & 0.118    & \textbf{0.299$^{*\dagger}$}  & 0.304$^{*\dagger}$  & 0.313$^{*\dagger}$  & 0.312$^*$\\
			ERM+idfW2V  & 0.106$^{*\dagger}$  & 0.114$^{*\dagger}$  & 0.121$^{*\dagger}$  & 0.118 & \textbf{0.299$^{*\dagger}$}  & 0.304$^{*\dagger}$  & \textbf{0.314$^{*\dagger}$}  & \textbf{0.314$^{*\dagger}$}  \\
			ERM+PV  & 0.103$^*$  & 0.115$^{*\dagger}$  & \textbf{0.122$^{*\dagger}$}  & \textbf{0.121$^*$} & \textbf{0.299$^{*\dagger}$}  & \textbf{0.307$^{*\dagger}$}  & \textbf{0.314$^{*\dagger}$}  & 0.313$^*$  \\
			ERM+PVC  & \textbf{0.107$^{*\dagger}$}  & 0.114$^{*\dagger}$  & 0.121$^{*\dagger}$  & 0.114  & 0.298$^{*\dagger}$  & 0.304$^{*\dagger}$  & 0.312$^*$  & 0.313$^{*\dagger}$  \\
			\hline
			\hline
			Distillation  & 0.099  & 0.104  & 0.109  & 0.111   & 0.292  & 0.299  & 0.311  & 0.313 \\
			\hline
			Distill+W2V  & \textbf{0.106$^*$} & \textbf{0.114$^*$}  & \textbf{0.120$^*$}  & 0.113 & 0.297$^*$  & 0.304$^*$  & 0.314$^*$  & 0.319$^*$   \\
			Distill+idfW2V  & \textbf{0.106$^*$}  & 0.113$^*$  & 0.116$^*$  & 0.115  & 0.297$^*$  & \textbf{0.306$^*$}  & 0.316$^*$  & 0.319$^*$   \\
			Distill+PV  & 0.103$^*$  & 0.110$^*$  & 0.118$^*$  & 0.116$^*$  & \textbf{0.298$^*$}  & \textbf{0.306$^*$}  & \textbf{0.317$^*$}  & \textbf{0.320$^*$}  \\
			Distill+PVC  & 0.105$^*$  & 0.112$^*$  & \textbf{0.120$^*$}  & \textbf{0.120$^*$}   & 0.297$^*$  & 0.304$^*$  & 0.315$^*$  & 0.317$^*$  \\
			\hline
		\end{tabular}  
	}
\end{table}
\begin{figure} [t]
	\centering
	\begin{subfigure}{.5\textwidth}
		\centering
		\includegraphics[width=2.5in]{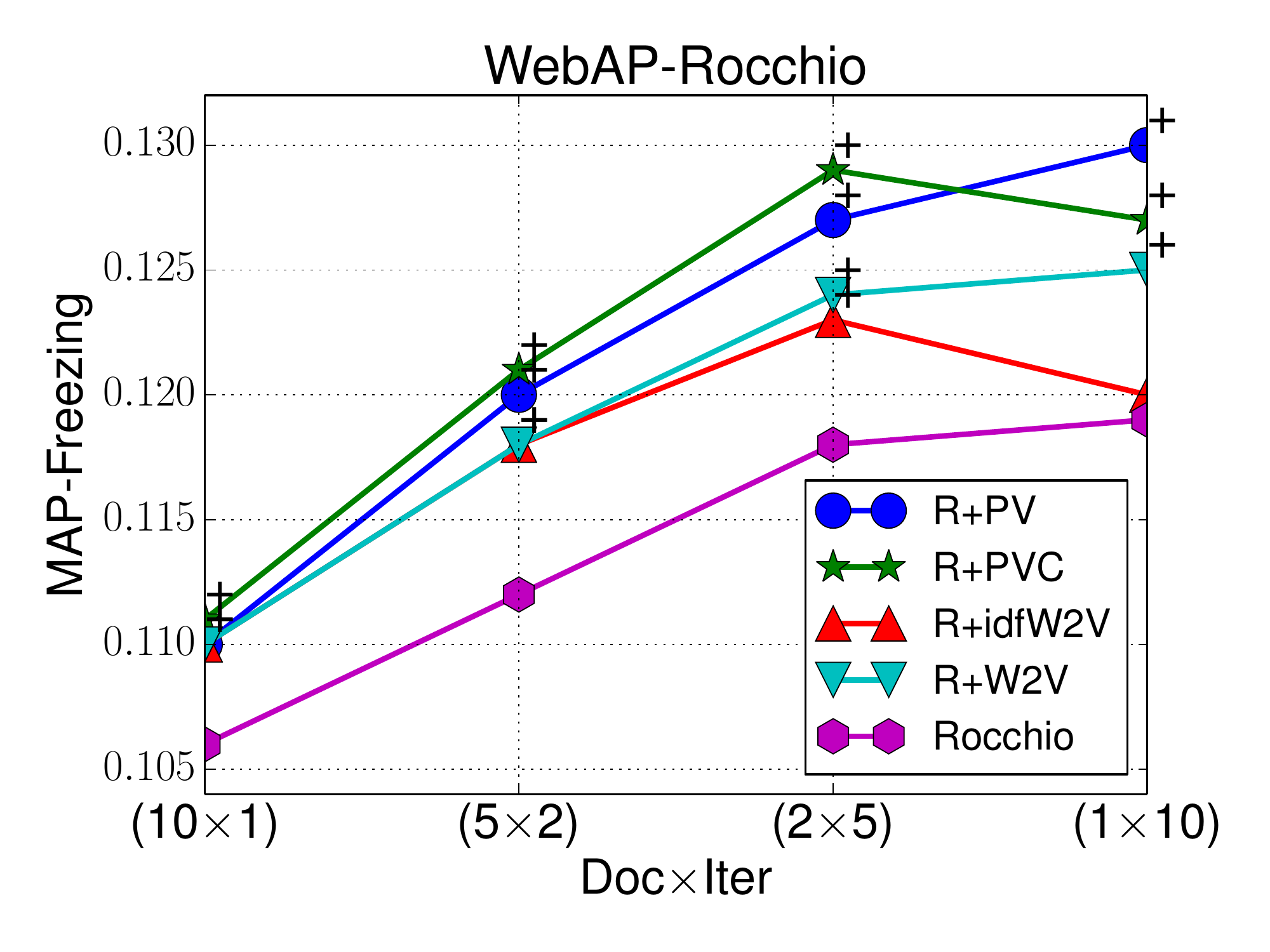}
		\label{fig:B}
	\end{subfigure}%
	\begin{subfigure}{.5\textwidth}
		\centering
		\includegraphics[width=2.5in]{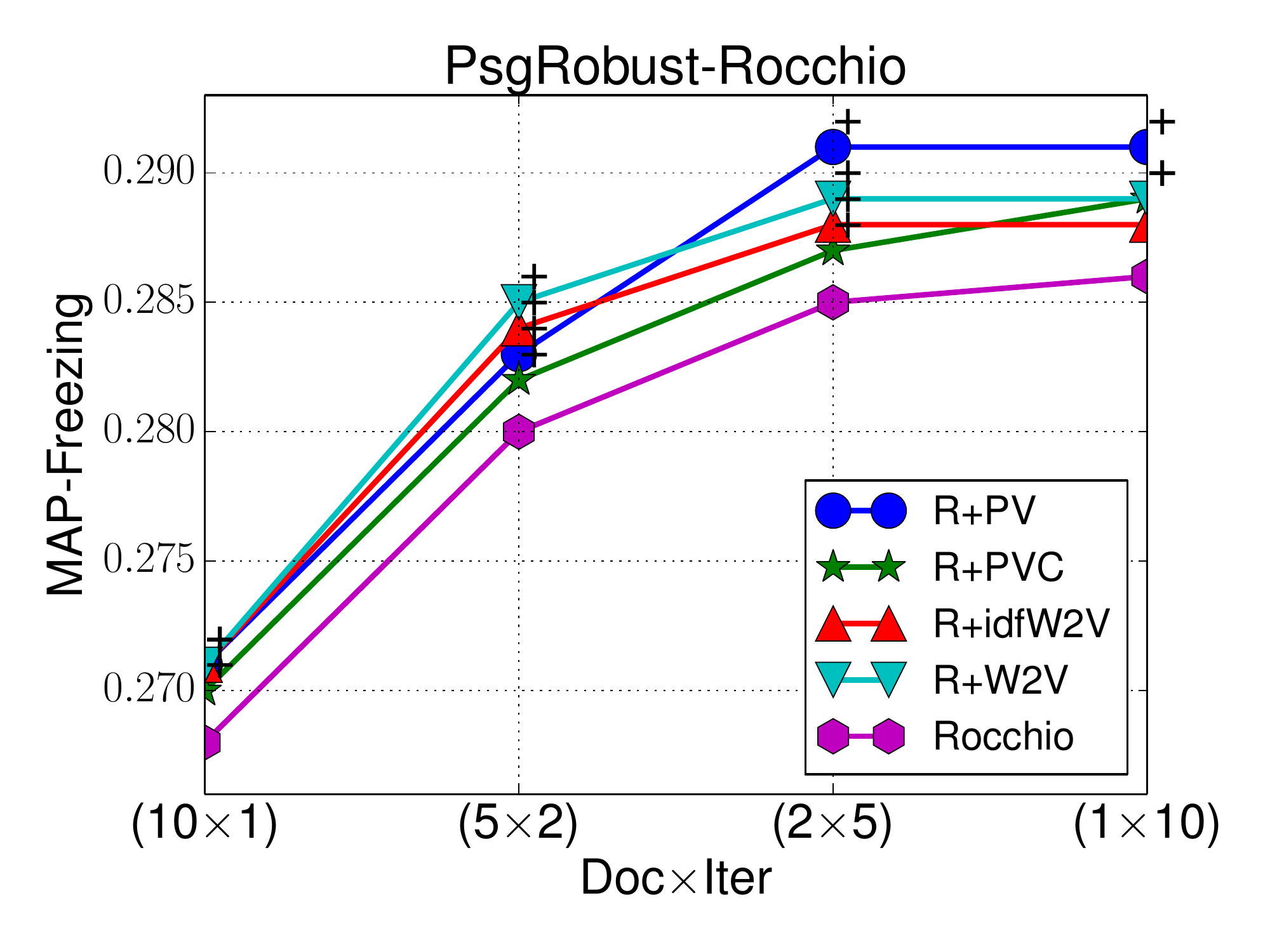}
		\label{fig:D}
	\end{subfigure}%
	\caption{Performance of our method with different paragraph representations compared with Rocchio. '+' means significant difference.
	}
	\label{fig:exp_IRF_EMB}
\end{figure}

\textbf{Results and Discussion. } 
We show the experimental results of using language model baselines (RM3, ERM, Distillation) in Table \ref{tab:embed_irf_psg} and include Rocchio as a baseline in Figure \ref{fig:exp_IRF_EMB}. We can see in general the four representations of paragraphs all can improve performance significantly over the word-based and embedding-based baselines under most iteration settings. ERM performs similar to RM3 on WebAP, and our method based on RM3 and ERM also perform similarly. On PsgRobust, ERM performs slightly better than RM3 and our method also performs slightly better combined with ERM than RM3. 
\footnote{The reason why ERM does not perform well will be shown in Section \ref{sec:exp_retrieve_given_onepsg} where we discuss the performance difference of ERM on the two tasks.}
This shows that incorporating passage-level semantic similarity in embedding space produces improvements to both the word-based RF models and the embedding-based RF model using semantic similarity at term level. 


The conclusion that IRF shows advantages over top-k feedback still holds when we incorporate word-based RF models with passage-level semantic match. In addition, there is no one representation better than the others all the time, which implies for different datasets, with different baselines, some representations show their advantages fitting the specific property underlying the setting. 

\subsection{Retrieval Given One Relevant Passage}
\label{sec:exp_retrieve_given_onepsg}
As we mentioned in Section \ref{sec:pv_irf_answer}, the small amount of text in answer passages during each iteration may not be enough to build word-based RF models.
The extreme case is when we have only one short passage as positive feedback.
Effective re-ranking after \textit{the first positive feedback} will show the user a second relevant answer in fewer iterations and make users less likely to leave after several interactions.
Therefore, it is particularly important to perform well given the first positive feedback from users.
We designed the second type of experiment to be answer retrieval given one relevant passage.

For each query, we randomly assign a relevant passage to the model as positive feedback and then retrieve 
from the remaining results.
To make the results more reliable, we randomly draw a relevant passage for each query ten times and do ten retrievals. 
Then we evaluate the performance of each model based on the overall rank lists from the ten retrievals. 
We take QL and BM25 as baseline retrieval models that do not consider feedback. 
Similar to the first group of experiments, we use RM3, Distillation, Rocchio as word-based RF baselines in the framework of LM and VSM, and ERM as the embedding-based RF baseline. 
We use $P@1$ (precision@1), $MRR$ (mean reciprocal rank) to evaluate the ability of a model to identify a second relevant passage in the next interaction given only one positive feedback. 
MAP at cutoff 100 measures the ability of the model to identify all the other relevant answers. 
\begin{table} [t]
	\caption{Performance of different IRF methods on finding other relevant answers given one relevant answer. '$*$' and '$\dagger$' denote significant improvements over word-based (RM3, Distillation, Rocchio) or embedding-based (ERM) baselines respectively.} 
	\label{tab:exp_emb_1rel}
	\centering
	\scalebox{0.9}{
		\begin{tabular}{l || l l l || l l l }
			\hline
			Dataset  & \multicolumn{3}{c ||}{\textbf{WebAP}} & \multicolumn{3}{c}{\textbf{PsgRobust}} \\
			\hline
			Model  & $P@1$ & $MRR$ & $MAP$  & $P@1$ & $MRR$ & $MAP$ \\
			\hline
			\hline
			QL & 0.259 & 0.373 & 0.071 & 0.367 & 0.486 & 0.231 \\
			\hline
			RM3 & 0.498 & 0.602 & 0.116 & 0.515 & 0.634 & 0.299 \\
			\hline
			ERM & 0.516 & 0.615 & 0.125 & 0.513 & 0.634 & 0.307 \\
			\hline
			RM3+W2V & 0.488 & 0.598 & 0.120$^*$ & 0.524$^{*\dagger}$ & 0.643$^{*\dagger}$ & \textbf{0.304$^*$} \\
			RM3+idfW2V & 0.488 & 0.597 & 0.120$^*$ & 0.525$^{*\dagger}$ & 0.643$^{*\dagger}$ & \textbf{0.304$^*$} \\
			RM3+PV & \textbf{0.525*} & 0.625$^*$ & 0.122$^*$ & 0.521 & 0.641$^*$ & 0.301$^*$ \\
			RM3+PVC & 0.524$^*$ & \textbf{0.635$^{*\dagger}$} & \textbf{0.123$^*$} & \textbf{0.526$^{*\dagger}$} & \textbf{0.644$^{*\dagger}$} & 0.303$^*$ \\
			\hline
			ERM+W2V & 0.513 & 0.622$^*$ & 0.131$^{*\dagger}$ & 0.529$^{*\dagger}$ & 0.648$^{*\dagger}$ & 0.312$^{*\dagger}$ \\
			ERM+idfW2V & 0.525$^*$ & 0.627$^*$ & 0.130$^{*\dagger}$ & \textbf{0.534$^{*\dagger}$} & 0.650$^{*\dagger}$ & 0.312$^{*\dagger}$ \\
			ERM+PV & \textbf{0.556$^{*\dagger}$} & 0.648$^{*\dagger}$ & 0.131$^{*\dagger}$ & 0.531$^{*\dagger}$ & 0.649$^{*\dagger}$ & 0.311$^{*\dagger}$ \\
			ERM+PVC & \textbf{0.556$^{*\dagger}$} & \textbf{0.658$^{*\dagger}$} & \textbf{0.134$^{*\dagger}$} & \textbf{0.534$^{*\dagger}$} & \textbf{0.653$^{*\dagger}$} & \textbf{0.313$^{*\dagger}$} \\
			\hline
			Distillation & 0.494 & 0.597 & 0.113 & 0.516 & 0.635 & 0.299 \\
			\hline
			Distill+W2V & 0.489 & 0.593 & 0.117$^*$  & \textbf{0.528}* & \textbf{0.645}* & \textbf{0.304}*  \\
			Distill+idfW2V & 0.489 & 0.595 & 0.117$^*$  & 0.525* & 0.643* & \textbf{0.304}*  \\
			Distill+PV & 0.519$^*$ & 0.621$^*$ & 0.120$^*$  & 0.514 & 0.638 & 0.297 \\
			Distill+PVC & \textbf{0.534}$^*$ & \textbf{0.638}$^*$ & \textbf{0.123}$^*$  & 0.524* & 0.643* & 0.303*  \\
			\hline
			\hline
			BM25 & 0.298 & 0.399 & 0.072 & 0.35 & 0.479 & 0.176 \\
			\hline
			Rocchio & 0.516 & 0.616 & 0.121 & 0.522 & 0.641 & 0.279 \\
			\hline
			Rocchio+W2V & 0.531 & 0.640$^*$ & 0.140$^*$  & 0.526 & 0.645* & \textbf{0.282}*  \\
			Rocchio+idfW2V & 0.536 & 0.642$^*$ & 0.139$^*$  & 0.526 & 0.644 & \textbf{0.282}*  \\
			Rocchio+PV & \textbf{0.576}$^*$ & \textbf{0.668}$^*$ & 0.138$^*$  & 0.518 & 0.642 & 0.280*  \\
			Rocchio+PVC & 0.560$^*$ & \textbf{0.668}$^*$ & \textbf{0.143}$^*$  & \textbf{0.528}* & \textbf{0.647}* & 0.281*  \\
			\hline
		\end{tabular}
	}
\end{table}

\textbf{Results and Discussion. }
In Table \ref{tab:exp_emb_1rel}, feedback methods are always better than their base retrieval models, i.e. QL, BM25.
In general, with the four paragraph representations, the improvements of $MAP$ over the baselines are always significant; $P@1$, $MRR$ can also be improved significantly in many cases. 
This shows that incorporating the passage semantic similarity can improve significantly over both the word-based RF baselines and the embedding-based RF baseline with only term-level semantic match information. 

In contrast to the IRF experiments, ERM performs much better than RM3 in this task. 
The reason may be that in the IRF experiments, there are more relevant passages for RM3 to extract expansion terms and alleviate the term mismatch problem, which makes the term-level semantic match from ERM less helpful. In this task, the text for RM3 is not enough to estimate an accurate model and ERM is effective with semantic match. 
Since our method considers semantic match at passage level, its benefit does not overlap with that from term-level semantic match.


On WebAP, our method combined with RM3 performs similarly to ERM when using PV and PVC and worse than ERM using W2V and idfW2V.  On PsgRobust, incorporating our method to RM3 performs better than ERM in terms of $P@1$ and $MRR$, but worse than ERM with $MAP$. This shows that incorporating embedding similarity to do RF at passage level or term level alone with little information are comparable to each other. When we combine these two ways of doing RF together, the performance can be further improved, which is shown from the significant improvements upon ERM when we add the passage similarity signal to ERM on both datasets. This is consistent with our claim that the semantic similarity of the passage level is complementary to the term level when combined with word-based RF models since they capture two different granularities of semantic match.  

Different from the IRF experiments, the performance of paragraph vectors are better than W2V and idfW2V. This indicates that when there is little feedback information, more accurate representations lead to better performance. In addition, with scarce user feedback, PVC is more effective than PV, probably because it is less susceptible to overfitting a small dataset due to many fewer parameters, i.e. vocabulary size versus corpus size.

\section{Conclusion and Future Work}
\label{sec:conclusion}

We first showed that IRF is effective on answer passage retrieval. 
Then we showed that, with passage-level semantic match 
the performance of iterative feedback and retrieval given one relevant passage can produce significant improvements compared with word-based RF models in the framework of both LM and VSM. The IRF experiments also show our method is better than the embedding-based baseline using term-level similarity. The retrieval experiment based on one relevance passage shows that combining the word and passage level granularities leads to the best performance.

Our method focuses more on user requests of ``more like this". We know diversity is also very important to provide users more informative results and we will take it into account in our future work. 
In addition, we will consider IRF on answer passage retrieval with end-to-end neural models.
\section{Acknowledgments}
This work was supported in part by the Center for Intelligent Information Retrieval and in part by NSF IIS-1715095. Any opinions, findings and conclusions or recommendations expressed in this material are those of the authors and do not necessarily reflect those of the sponsor.

\bibliographystyle{splncs04}
\bibliography{ECIR2019-answerIRF} 

\end{document}